\documentclass{bmcart}
\usepackage{cite}

\usepackage[utf8]{inputenc} 
\usepackage{amsfonts}
\usepackage{amssymb}
\usepackage{color}
\usepackage{amsmath}
\usepackage{graphicx}
\usepackage{txfonts}
\startlocaldefs
   \newcommand{\Ham}{\mathcal H}
   \newcommand{\ket}[1]{\left\vert#1\right\rangle}
   \newcommand{\bra}[1]{\left\langle#1\right\vert}
\endlocaldefs

\begin{document}

\begin{frontmatter}

\begin{fmbox}
\dochead{Research}

\title{Phase diagram of a QED-cavity array coupled via a $N$-type level scheme}

\author[
   addressref={aff1},
   email={jiasen.jin@sns.it}
]{\inits{J}\fnm{Jiasen} \snm{Jin}}
\author[
   addressref={aff1,aff2},
   email={fazio@sns.it}
]{\inits{R}\fnm{Rosario} \snm{Fazio}}
\author[
   addressref={aff1},
   email={rossini@sns.it}
]{\inits{D}\fnm{Davide} \snm{Rossini}}

\address[id=aff1]{
  \orgname{NEST, Scuola Normale Superiore and Istituto di Nanoscienze - CNR},
  \street{Piazza dei Cavalieri, 7},
  \postcode{56126}
  \city{Pisa},
  \cny{Italy}
}

\address[id=aff2]{
  \orgname{Center for Quantum Technologies, National University of Singapore},
  \postcode{117543}
  \city{Singapore}
}

\end{fmbox}

\begin{abstractbox}

\begin{abstract}
We study the zero-temperature phase diagram of a one-dimensional array of QED cavities where,
besides the single-photon hopping, an additional coupling between neighboring cavities is mediated
by an $N$-type four-level system. By varying the relative strength of the various couplings,
the array is shown to exhibit a variety of quantum phases including a polaritonic Mott insulator, a
density-wave and a superfluid phase. Our results have been obtained by means of numerical density-matrix
renormalization group calculations. The phase diagram was obtained by analyzing the energy gaps for the
polaritons, as well as through a study of two-point correlation functions.
\end{abstract}



\end{abstractbox}


\end{frontmatter}


\section{Introduction}

The recent impressive advances in the field of quantum simulators
allowed to probe the many-body physics of strongly correlated systems
at the level of the single quantum object. At present cold atoms trapped in optical lattices
can be considered among the most promising examples of quantum simulators.
By means of ultracold atomic and molecular gases, it is nowadays possible
to reach a degree of control and accuracy in engineering the dynamics of many-body
systems that were unimaginable in previously.
As a consequence, the coherent quantum dynamics emerging from carefully tailored
microscopic Hamiltonians can now be tested experimentally~\cite{Bloch2008}.
It has been possible, just to recall one example, to implement the Bose-Hubbard (BH)
model~\cite{Fisher1989, Jaksch1998} and to detect its zero-temperature superfluid (SF)
to Mott insulator (MI) quantum phase transition~\cite{Greiner2002}.
Other models involving spinor gases, Fermi systems, Bose-Fermi mixtures,
or dipolar gases have been also devised and realized, providing an even
richer phase diagram (see for example the review~\cite{Lewenstein2007}).
We mention the stabilization of density-wave (DW) phases for bosons,
as well as more peculiar topological or supersolid orderings, which can
arise in the presence of finite-range interactions~\cite{Pfau2009}.

More recently a novel kind of many-body quantum simulator has been introduced,
based on the idea to use single photons as quantum objects.
Since photons hardly interact in open space, the most natural way to significantly
increase their interactions is to trap them into an optical QED cavity, and couple
the field with atoms/molecules inside it in order to create an optical nonlinearity.
If the nonlinearity is sufficiently large, the so called photon blockade
sets in~\cite{Schmidt1996, Imamoglu1997}, namely, the presence of a single photon inside
a cavity prevents a second one to enter it.
In the rotating-wave approximation, the simplest light-matter interaction scheme
of this type can be accurately described by the Jaynes-Cummings (JC) model.
By arranging an array of cavities coupled through the photon hopping,
such to generate a competition between the hopping and the on-site nonlinearities,
one can devise a setup that is well described by the so called
Jaynes-Cummings-Hubbard (JCH) model~\cite{Hartmann2006, Greentree2006, Angelakis2007}.

In many respects, if one ignores dissipation, the physics emerging from the JCH Hamiltonian
resembles, at low-energies, that of an effective BH model.
Probably the main difference between the two systems is
that, instead of having neutral bosons as building blocks of the model, in the JCH Hamiltonian
one has to think in terms of polaritons, {\it i.e.}, combined photonic/atomic excitations.
Many different works already addressed the JCH equilibrium
phase diagram with analytical, as well as numerical methods, leading to a fairly complete theoretical
understanding of the nature and the location of the emerging
phases and phase transitions in terms of the parameters governing the system
(the field has been recently reviewed in, {\it e.g.}, Refs.~\cite{Review1, Review2, Review3, Review4}).

Additional interest in cavity arrays comes from the fact that these systems can be naturally
considered as open-system quantum simulators. Some related features have been recently
explored~\cite{NonEq1,NonEq2,NonEq3,NonEq4,NonEq5,NonEq6,NonEq7,Jin2013,Jin2014}.
In the following we will not touch on this and consider only the ``equilibrium'' phase diagram.

This intense activity has been very recently boosted by the first experiments on
QED cavity arrays~\cite{Underwood2012, Abbarchi2013, Toyoda2013}. As of today,
the most concrete possibility to realize controllable and scalable quantum simulators with
cavity arrays involves circuit-QED cavities~\cite{Lucero2012, Steffen2013, Chen2014}.

So far the coupling between cavities
has been mostly considered through photon hopping.
Only few works started addressing more general schemes,
where the cavity coupling can be induced also by means of non-linear
elements~\cite{Zueco2012, Peropadre2013, Jin2013, Jin2014}.
Such configurations include cross-Kerr interactions and/or
correlated hopping terms, which lead to generalizations of the JCH model
in a way similar to the extended BH (EBH) Hamiltonian
for atoms with large dipole momentum loaded in optical lattices~\cite{Sowinski2012}.
The underlying physical model is believed to possess a much richer structure,
with the emergence of exotic phases of correlated polaritons.
It is particularly interesting to address these schemes in one-dimensional (1D) systems,
where interactions become crucial to stabilize exotic phases
of matter~\cite{DallaTorre2006A, DallaTorre2006B, Sowinski2012, Rossini2012, Deng2013}.
These notably include a series of nontrivial density-wave (DW) states,
which can arise in the strong coupling regime~\cite{Wikberg2012},
as well as supersolidity and phase-separation effects~\cite{Batrouni2013, Batrouni2014}.
Extension to consider also counter-rotating terms in the ultrastrong coupling
regime, thus leading to the so called Rabi-Hubbard model~\cite{Schiro2012},
have been investigated~\cite{Kumar2013}.
However we are not aware of numerical investigations of coupled cavity models
beyond the JCH and Rabi-Hubbard model.

In all such situations, non-perturbative, either numerical or analytical calculations are necessary.
Here the density-matrix renormalization group (DMRG) algorithm~\cite{dmrgA, dmrgB}
has been employed to work out the quantitative zero-temperature phase diagram
of the JCH model~\cite{Rossini2007, Rossini2008, Souza2013}.
This is a particularly efficient method for the statics of 1D many-body problems.
Its key strategy consists in constructing a portion of the system (called block) and then
recursively enlarge it. At each step, the basis of the corresponding Hamiltonian is truncated
to a given value $m$, so that one can manage the Hamiltonian in an effective Hilbert space
of fixed dimensions, as the physical system grows. This truncation is performed
by retaining the eigenstates corresponding to the $m$ highest eigenvalues
of the reduced density matrix of the block.

The aim of this paper is to quantitatively study a generalization
of the JCH Hamiltonian, aimed at taking into account an effective
nearest-neighbor nonlinearity between cavities mediated
by an $N$-type four-level system as discussed for two cavities in Ref.~\cite{Hu2011}.
The presence of this coupling leads to an effective cross-Kerr non-linearity.
An analysis at the mean-field level of a dissipative open EBH as an effective model for nonlinearly
coupled cavities has been performed, unveiling the emergence of novel
photon crystal and supersolid phases~\cite{Jin2013, Jin2014}.
Here we do not resort to the effective EBH model and analyze the full model as introduced
in~\cite{Hu2011}. Using the DMRG algorithm, we work out the 1D ground-state phase diagram.
We show that a physics similar to the EBH model appears,
with a rich phase diagram including gapless SF, as well as MI
and DW phases of polaritons.
We postpone the analysis of the interplay of driving and dissipation to a future work.

The paper is organized as follows.
In the next two sections we introduce the model of coupled cavities
of our interest (Sec.~\ref{sec:model}) and the quantities we are going
to address, namely the energy gaps, and the staggered
number-number correlations (Sec.~\ref{sec:quantities}).
In Sec.~\ref{sec:pd} we discuss the zero-temperature equilibrium phase diagram,
focusing on the MI/SF boundary and on the boundary separating the DW from the other phases.
Finally, in Sec.~\ref{sec:summary} we draw our conclusions.

\section{The model}
\label{sec:model}

\begin{figure}[!b]
  \includegraphics[width=0.9\hsize]{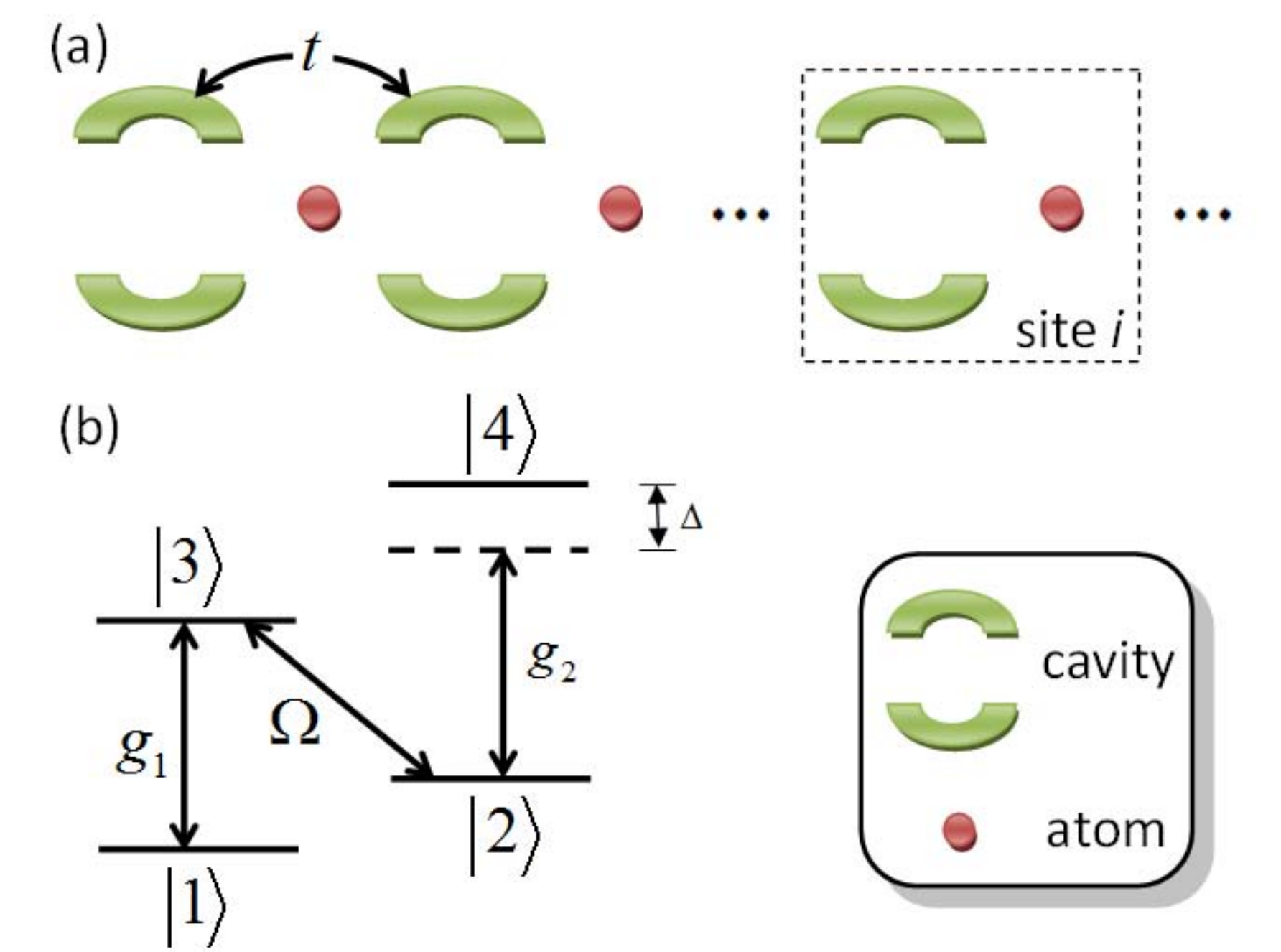}
  \caption{{\bf Scheme for nonlinearly coupled QED cavities.}
    (a) An array of QED cavities nonlinearly coupled by $N$-type atoms.
    The photon hopping between nearest-neighbor cavities has a strength $t$.
    Each effective site is composed of a cavity and an atom (dashed box).
    (b) Level structure of the $N$-type atoms.
    The transition $\ket{1} \leftrightarrow \ket{3}$ is resonantly coupled
    to the cavity mode of its own site with strength $g_1$,
    while $\ket{2} \leftrightarrow \ket{4}$ is coupled to the cavity mode of
    its right nearest-neighbor site with strength $g_2$, and has a detuning $\Delta$.
    The transition $\ket{2} \leftrightarrow \ket{3}$ is in resonance with
    an external laser field of strength $\Omega$.}
  \label{fig1}
\end{figure}

Let us consider a 1D array of QED cavities,
where photons can hop between neighboring cavities.
Moreover two adjacent resonators are also nonlinearly coupled to each other
via a $N$-type four-level system, as shown in Fig.~\ref{fig1}(a).
For the sake of clarity in our description, we shall divide the 1D array
into coupled effective sites composed of a cavity and an atom.
The four levels are denoted by
$\{ \ket{i} \}_{i = 1 \ldots 4}$, and are depicted in Fig.~\ref{fig1}(b).
An external laser with frequency $\Omega$ resonantly drives
the transition $\ket{3} \leftrightarrow \ket{2}$.
The transition $\ket{1} \leftrightarrow \ket{3}$
is resonantly coupled to the cavity mode of the same site with strength $g_1$,
while the transition $\ket{2} \leftrightarrow \ket{4}$ couples to the cavity mode
of its right nearest-neighbor site with strength $g_2$, and a detuning $\Delta$.

The use of such $N$-type atom for generating large Kerr nonlinearity
has been extensively studied in the literature~\cite{Schmidt1996,Imamoglu1997,Rebic2009},
however the vast majority of the scenarios only focused on a single-mode cavity.
Our work is inspired by the idea of Ref.~\cite{Hu2011}, where the cross-Kerr nonlinearity
is generated between two different and neighboring cavities, in circuit-QED systems.
In practice, we use the unbalanced couplings of atomic transition $|1\rangle\leftrightarrow|3\rangle$
with left cavity mode, and $|2\rangle\leftrightarrow|4\rangle$ with right cavity mode respectively,
in order to generate the local ($g_1$) and nonlocal ($g_2$) nonlinearities of our many-body system.
This kind of four-level artificial molecule can be realized using two Josephson
transmon qubits coupled by a superconducting quantum interference device.

Using the interaction picture and in the rotating-wave approximation, the system Hamiltonian reads
\begin{equation}
  \Ham = \sum_i \bigg[ \Delta \sigma^{44}_{i} +
    \left( \Omega \sigma^{23}_i + g_1 \sigma_i^{13} a_i^\dagger + \mathrm{H.c.} \right)
    + \left(-t a_{i} \, a_{i+1}^\dagger + g_2 \sigma_i^{24} a_{i+1}^\dagger
    + \mathrm{H.c.} \right) \bigg] \,,
  \label{hamiltonian}
\end{equation}
where $\sigma^{mn} = \ket{m} \bra{n}, (m,n = 1,2,3,4)$,
and $a$ $(a^\dagger)$ is the annihilation (creation) operator of the cavity mode.
The subscripts denote the site position along the 1D chain.
The first three terms in the r.h.s. of Eq.~\eqref{hamiltonian} describe the local
terms and the nonlinearities on each site.
Inside the latter brackets, the first term is the photon hopping, while the second term
describes the coupling of the atom to its right neighboring cavity, which generates
an effective nonlocal cross-Kerr nonlinearity between the two cavities.

Hereafter we concentrate on the 1D model in Eq.~\eqref{hamiltonian}
at zero temperature, specifically addressing the case without dissipation with DMRG.
Let us also fix the Hamiltonian quantities in units of $\Omega$, set $\hbar = 1$,
and work with open boundary conditions.
We recall that, in the presence of dissipation, the problem becomes much more difficult
to be handled numerically~\footnote{It is however possible to address the effect
of dissipation with a DMRG approach in a 1D chain, when this is described
by a master equation within the Lindblad formalism.
In the language of tensor networks, one has to generalize the matrix-product-state
ansatz to a matrix-product-density-operator ansatz for mixed states,
as originally proposed in Refs.~\cite{Verstraete2004, Zwolak2004}.
The computational complexity is greater than for static computations, and
is eventually related to the amount of entanglement in the steady state.}.

For the system we are considering here, in the strong coupling
regime atoms and photons cannot be considered as two separate entities.
It is thus natural to investigate the phase diagram in terms of combined
atomic/photonic modes, named polaritons.
The polaritonic number operator on each site $i$, representing the number
of local excitations, is defined as
$n^{\mathrm{pol}}_i=2\sigma_i^{44}+\sigma^{33}_i+\sigma^{22}_i+a^\dagger_ia_i$.
For the closed system described by the Hamiltonian~\eqref{hamiltonian},
the total number $N^{\rm pol} = \sum_i n^{\rm pol}_i$ of such polaritons
is a conserved quantity.
In the following we work in the canonical ensemble for polaritons,
and focus on the integer filling situation.

\section{Energy gaps and correlation functions}
\label{sec:quantities}

The different nature of the various phases is sensitive to a number
of properties which we are going to focus on.
Here we are going to study quantities that resemble those characterizing
the various phases of the EBH model~\cite{Rossini2012}.

First of all, the ground-state energy gap is an important indicator
which characterizes the presence or absence of criticality in the model.
In particular, in the critical SF phase, the {\it charge gap} vanishes
in the thermodynamic limit. On the other side, in the insulating
MI and DW phases, such gap remains finite. In order to make connection with a similar
notation in the EBH model, below we introduce the so called charge and neutral gaps
referring respectively to the gaps corresponding to adding one extra particle (``charge" sector) or
remaining with the same number of particles (``neutral" sector). We stress however that in the
present model the excitation carry no real charge. This has to be understood only as a
convention.

The charge gap is defined as
\begin{equation}
  \Delta E_c = \Delta E^+ - \Delta E^- \, ,
\end{equation}
where, in the canonical ensemble, $\Delta E^+$ ($\Delta E^-$) denotes
the extra energy needed to add (remove) one particle,
{\it i.e.} one polariton, in the system.
In the specific, focusing on the unit filling,
$\Delta E^+ = E_{L+1}-E_{L}$ and $\Delta E^- = E_{L}-E_{L-1}$,
where $E_L$ is the ground-state energy per site of an $L$-sites
cavity-array with exactly $L$ excitations,
and $E_{L+1}$ ($E_{L-1}$) is the corresponding energy per site
with one excitation more (less).
It is therefore possible to extrapolate $\Delta E_c$ by running
three different DMRG simulations with fixed number of polaritons
$N^{\rm pol} = L-1, \, L, \, L+1$~\cite{KuhnerA, KuhnerB}.

While the charge gap is able to detect particle-hole excitations,
in some circumstances it is possible that the dominant
low-energy excitations are of a different type.
Their presence can be detected only by the so called {\it neutral gap}
at a fixed number of particles,
\begin{equation}
  \Delta E_n = E_L^{1} - E_L \, ,
\end{equation}
where, again working in the canonical ensemble, $E_{L}^{1}$
denotes the first excited energy per site
of an $L$-site system with $L$ excitations.

In the following, we also focus on the analysis of the staggered diagonal order
for the polaritons, in order to distinguish the DW from the other phases.
We do this by investigating the two-point correlation function
\begin{equation}
  C_{\mathrm{DW}}(r)=(-1)^r \langle \delta n_i^{\mathrm{pol}} \delta n_{i+r}^{\mathrm{pol}} \rangle \, ,
  \label{eq:DWcorr}
\end{equation}
where $\delta n_i^{\mathrm{pol}} = n^{\mathrm{pol}}_i-\bar{n}$ denotes the polariton fluctuation
from the average filling $\bar{n}$.
The order parameter identifying the DW phase is thus given by:
$\mathcal{O}_{\mathrm{DW}}\equiv\lim_{r\rightarrow\infty}{C_{\mathrm{DW}}(r)}$.
A finite value of $\mathcal{O}_{\mathrm{DW}}$ indicates a tendency to establish,
in the thermodynamic limit, a staggered occupation of the polaritons.
On the other side, in the MI as well as the SF phases,
$C_{\mathrm{DW}}(r)$ vanishes exponentially with increasing distance $r$.

\section{Phase diagram}
\label{sec:pd}

\begin{figure}[!t]
  \includegraphics[width = 1\hsize]{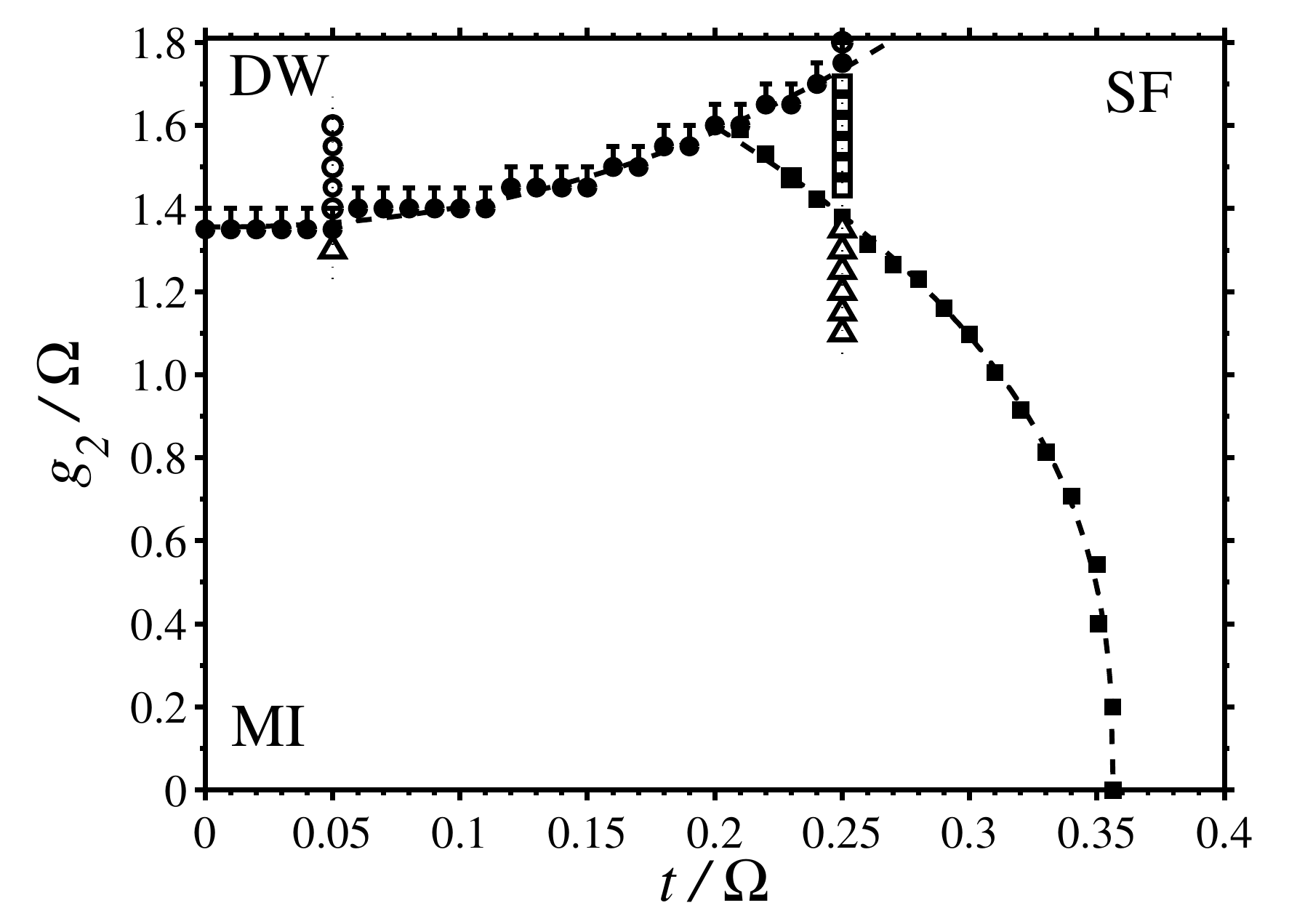}
  \caption{{\bf Phase diagram.}
    Ground-state phase diagram for a 1D system of coupled cavities
    described by Hamiltonian~\eqref{hamiltonian} in the $g_2-t$ parameter space.
    Here and in the subsequent figures
    we choose $g_1 / \Omega = 0.8$, $\Delta / \Omega = -2$.
    The various symbols correspond to points belonging to different phases
    (triangles = MI, circles = DW, squares = SF).
    Filled points lie close to a phase transition, and are used to
    draw the interpolating curves (dashed lines).
    The DW-to-MI/SF boundary (filled circles) has been obtained
    by analyzing the density-wave order parameter,
    while the MI-SF boundary (filled squares) through the charge gap.
    The two vertical dotted lines denoted two cuts in the phase diagram which
    will be analyzed in details below.
    The error bars in the points characterizing the DW-to-MI/SF
    boundary take into account the discretization of the $g_2$ values
    that we adopted in our numerical simulations (see the text).}
  \label{fig2}
\end{figure}

The zero-temperature phase diagram of model~\eqref{hamiltonian}, at unit polariton
filling $\bar{n} = 1$ and in the $g_2-t$ plane, is summarized in Fig.~\ref{fig2}.
We observe that three different phases can be stabilized.
Their boundaries have been obtained by means of a finite-size scaling
of the numerical data, for systems up to $L=300$ sites.
In our simulations we imposed a cutoff photon number in each cavity,
such that $n^{\rm phot}_i \leq 3$.
We also truncated the effective Hilbert space dimension to a value
$m=80$ in all the simulations, except for those shown in Fig.~\ref{fig5}
for the neutral energy gap (see the discussion in Sec.~\ref{sec:neutral_gap}).
We checked that, by increasing $m$ and the local fock-space truncation
over the photon number, the results concerning the charge gap
and the DW order parameter do not change on the scales shown here.

For small photon hopping ($t / \Omega \lessapprox 0.2$),
by increasing the nonlocal nonlinearity $g_2$ the system exhibits
a direct transition from the MI to the DW phase.
On the other hand, for $t / \Omega \gtrapprox 0.2$, the MI-to-DW transition
is mediated by an extended region appearing at intermediate $g_2$ values,
where the system stabilizes into a gapless SF.
In the following we are going to elucidate our finite-size scaling procedure
and how we were able to distinguish between the different phases.

\subsection{Boundary between MI and SF phases}

In the limit of small $g_2$ and $t$ values, the dominant presence
of the on-site interactions stabilize the system into a MI phase
with exactly one polariton per cavity ($\bar{n} = 1$),
and where the charge energy gap has a finite value.
As long as the hopping strength $t$ is progressively increased
(and for fixed $g_1, \, g_2$), the system eventually enters a SF phase,
with a vanishing gap.
The filled squares of Fig.~\ref{fig2} denoting the MI/SF boundaries
have been obtained by means of a finite-size scaling of the charge gap.
We performed simulations up to $L=100$ sites and analyzed whether the gap
closes or remains finite in the thermodynamic limit $L \to \infty$.

\begin{figure}[!t]
  \includegraphics[width = 0.495\hsize]{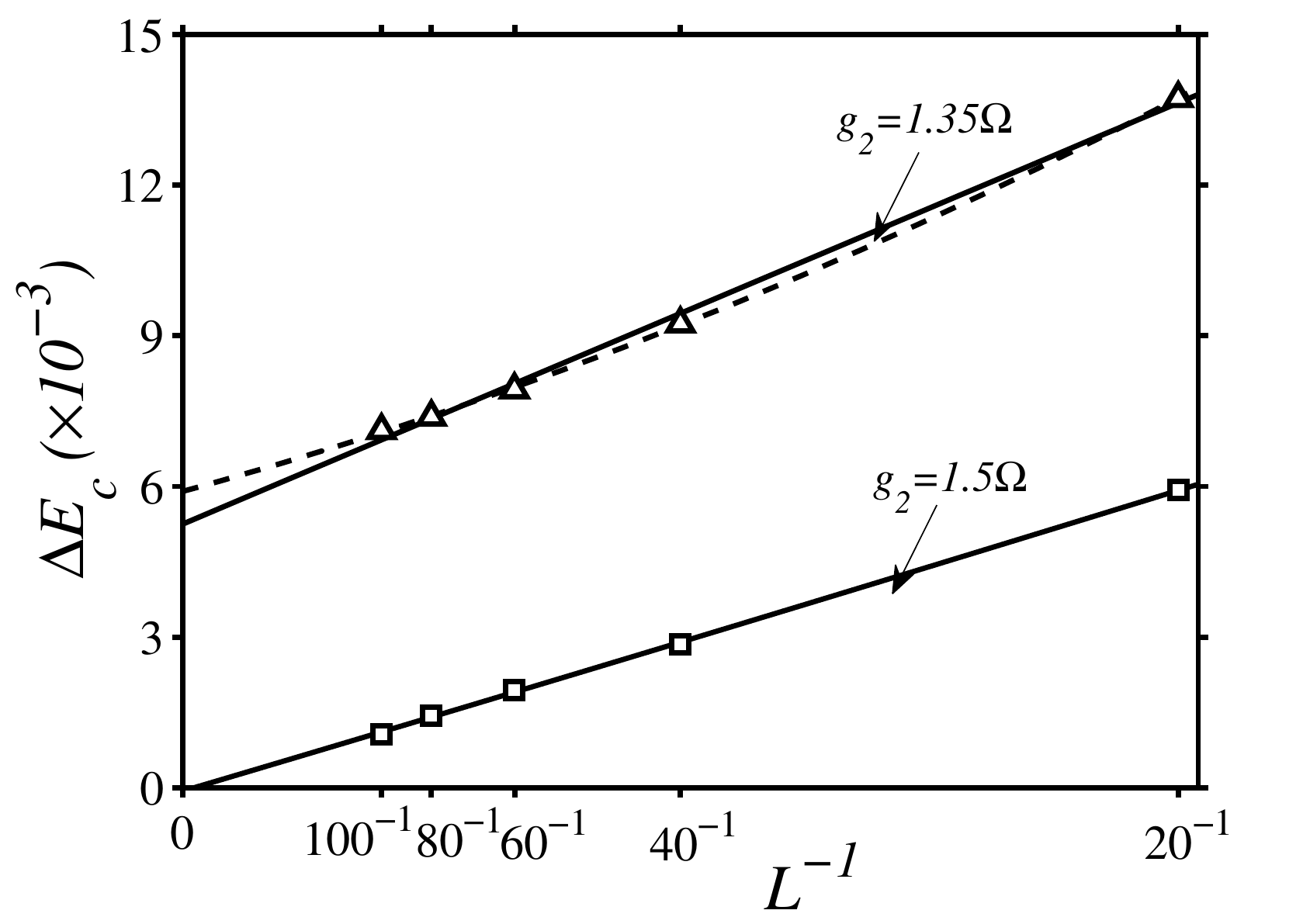}
  \includegraphics[width = 0.495\hsize]{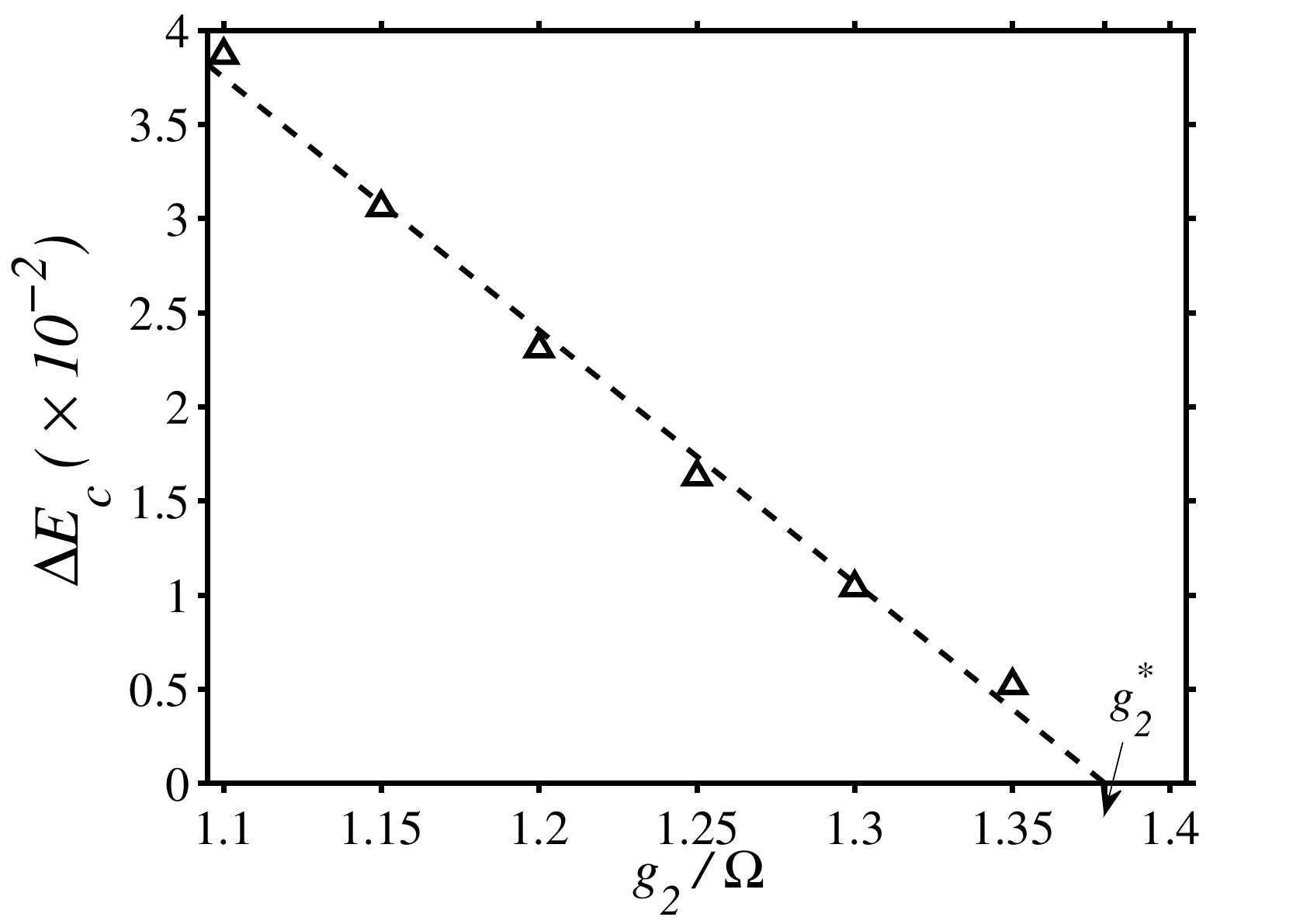}
  \caption{{\bf Analysis of the MI-SF Boundary.}
    Left panel: system-size dependence of the charge gap $\Delta E_c$ per site,
    for fixed $t/\Omega = 0.25$ and two values of $g_2$ in the MI ($g_2/\Omega = 1.35$)
    and in the SF ($g_2/\Omega = 1.5$) phase. Symbols denote the DMRG results.
    Solid and dashed lines are linear and quadratic fitting curves, respectively.
    The difference between the extrapolated values in the two fits
    $\Delta E_c^{\infty} = \lim_{L \to \infty} \Delta E_c^{(L)}$
    is negligible.
    Right panel: determination of the critical $g_2^*$ value
    for the quantum phase transition. 
    The triangles denote the charge gap per site $\Delta E_c^{\infty}$
    at the thermodynamic limit, as extrapolated in the left panel.
    The dashed line is a best linear fit of the data {\it vs.} $g_2$.
    The critical point is obtained when $\Delta E_c$ vanishes.
    For $t / \Omega =0.25$, we get $g_2^* / \Omega \approx 1.379$.}
  \label{fig3}
\end{figure}

In Fig.~\ref{fig3}, left panel, we highlight the size-dependence of $\Delta E_c$
as a function of $1/L$ for two points in the phase space
close to the MI/SF transition (see points along the dotted line in Fig.~\ref{fig2}).
We expect to see a quadratic dependence $\Delta E_c \sim L^{-2}$
(dashed line) at large $L$~\cite{KuhnerA, KuhnerB}, however a linear extrapolation
(solid line) is already a good approximation to the scaling,
and we can use it to determine $\Delta E_c$ in the thermodynamics limit.
Indeed, we observe that the difference between quadratic
and linear extrapolation is tiny ($\lesssim 10^{-3}$) and does not produce any
distinguishable modification on the scale of Fig.~\ref{fig2}.
In the specific case of Fig.~\ref{fig3}, we fixed $t / \Omega = 0.25$
and chose two different values of $g_2 / \Omega$ corresponding to configurations
in the gapped MI ($g_2 / \Omega = 1.35$, triangles) and in the gapless SF phase
($g_2 / \Omega = 1.5$, squares).
The MI is signaled by an extrapolated finite value of $\lim_{L \to \infty} \Delta E_c > 0$,
while in the SF this is zero.

In order to locate the critical $g_2$ for a given value of $t$
(filled squares in Fig.~\ref{fig2}) we perform a linear extrapolation
of the charge gaps in the vicinity of the critical value of $g_2$.
An example of such procedure is shown in the right panel of Fig.~\ref{fig3},
where we plot $\Delta E_c$ as a function of $g_2$, when this is close
to the phase transition (in the specific, here we set $t / \Omega = 0.25$).
After a linear extrapolation, we get a critical $g_2$ value
corresponding to $g_2^* / \Omega \approx 1.379$.
An analogous procedure is repeated for all the filled squares
shown in Fig.~\ref{fig2}, thus identifying the MI/SF boundary.

\subsection{Boundary of the DW phase}

\begin{figure}[!t]
  \includegraphics[width = 0.495\hsize]{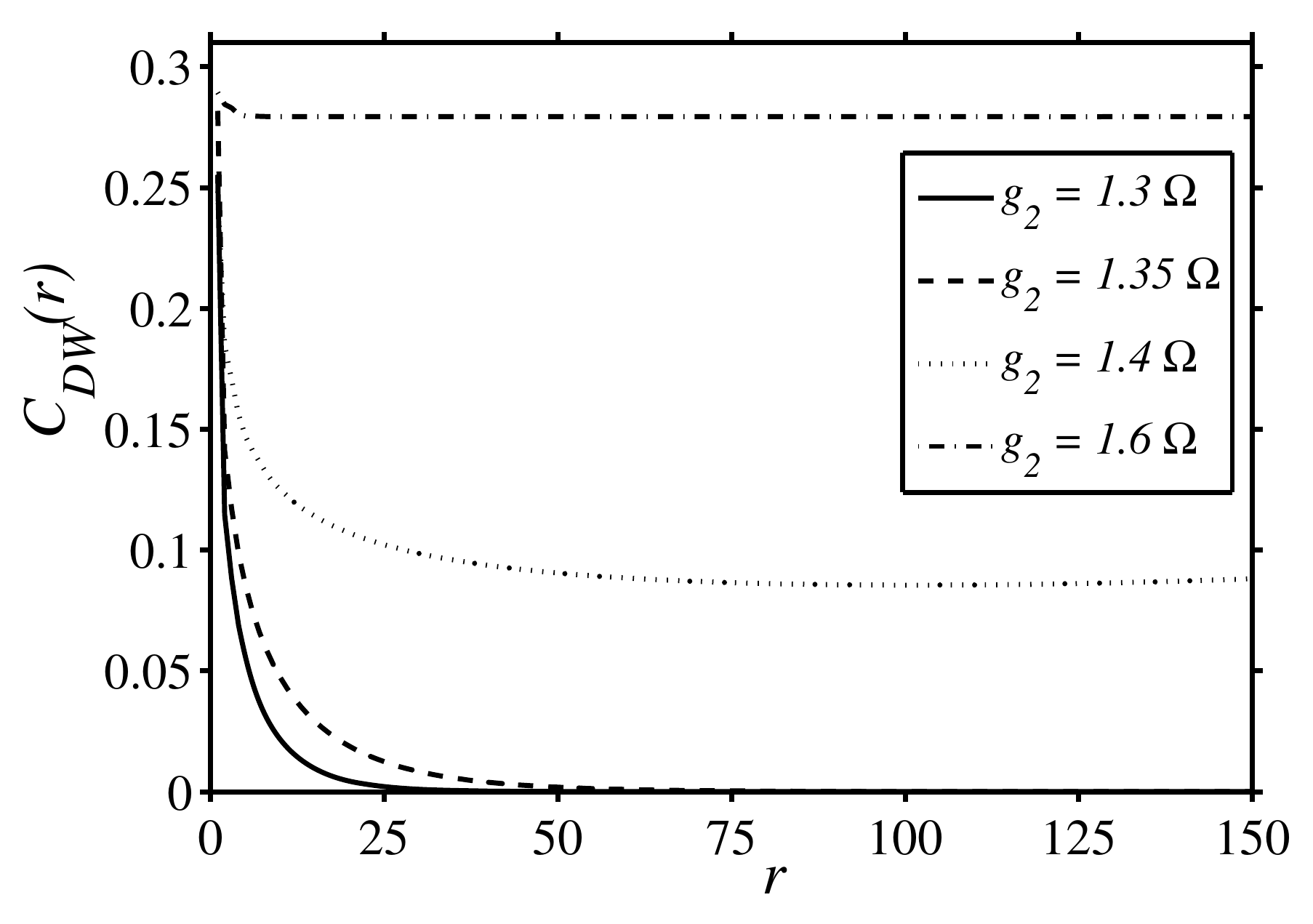}
  \includegraphics[width = 0.495\hsize]{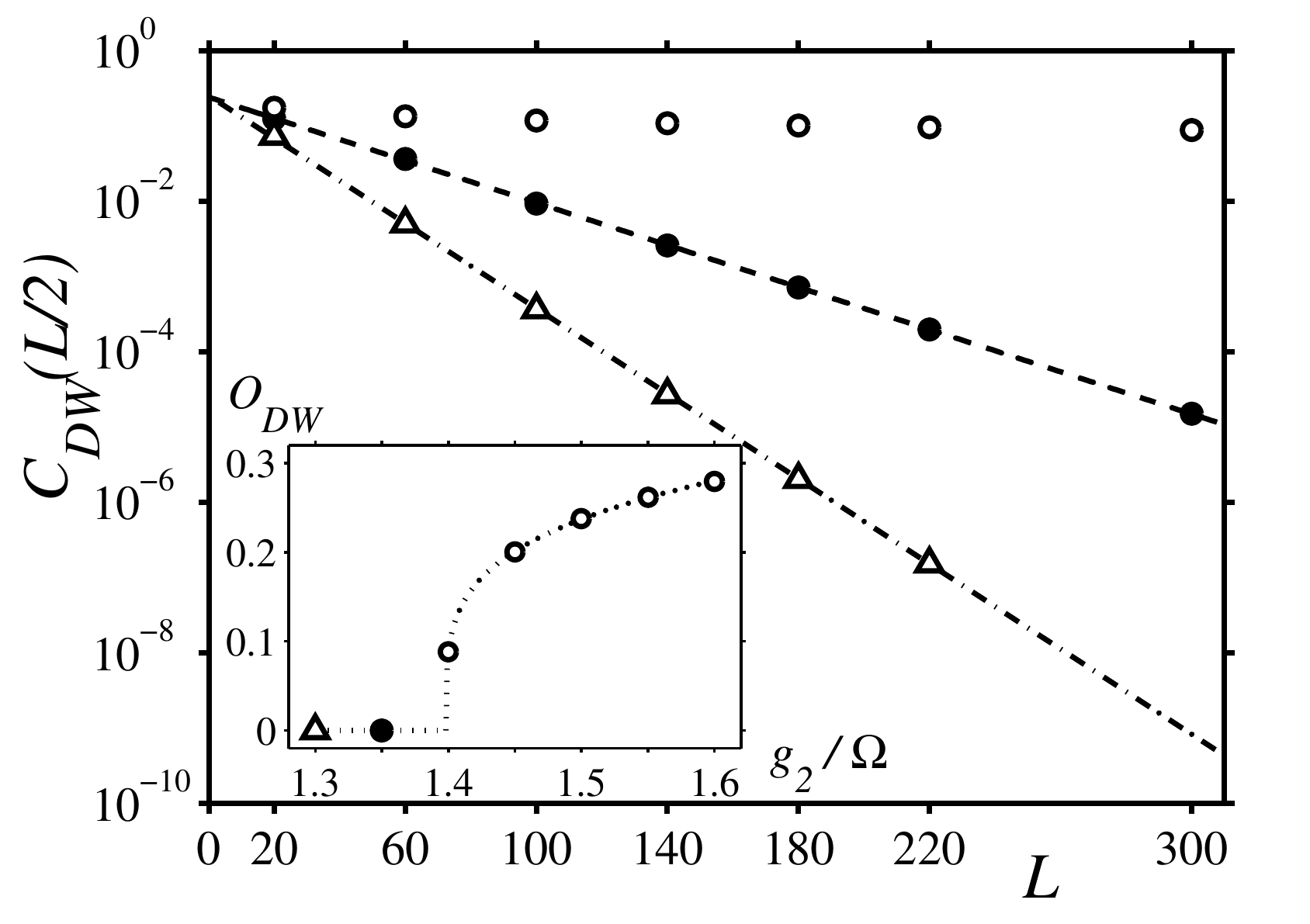}
  \caption{{\bf Determination of the DW boundaries.}
    The two-point correlation function $C_{\mathrm{DW}}(r)$ for the polariton
    number and its asymptotic value near the MI-DW quantum phase transition.
    Here we fix $t/\Omega = 0.05$ and vary $g_2/\Omega$.
    Left panel: behavior at fixed system size $L=300$, as a function
    of the distance $r$ and for different values of
    $g_2/\Omega = 1.3$, $1.35$, $1.4$, and $1.6$.
    To minimize boundary effects, we chose the two points $(i, i+r)$
    symmetrically with respect to the center of the array.
    Right panel: finite-size scaling close to the transition.
    Empty circles, filled circles, and triangles respectively are for
    $g_2 / \Omega = 1.4$ (DW), $1.35$ (near the critical point), and $1.3$ (MI phase).
    In the MI phase, $C_{\mathrm{DW}}$ vanishes exponentially with $L$,
    according to the fits: $C_{\mathrm{DW}}^{g_2=1.35} \approx 0.241 \times e^{-0.032 L}$
    and $C_{\mathrm{DW}}^{g_2=1.3} \approx 0.253 \times e^{-0.065 L}$.
    In the DW phase, $C_{\mathrm{DW}}$ converges to a finite value.
    The inset shows such obtained asymptotic value $\mathcal{O}_{\mathrm{DW}}$,
    as a function of $g_2$.}
  \label{fig4}
\end{figure}

The DW phase is characterized by a finite order parameter ${\mathcal O}_{\mathrm{DW}}$.
Let us therefore look at the two-point staggered correlator in Eq.~\eqref{eq:DWcorr}.
Since in DMRG simulations we are employing open boundary conditions,
to minimize the border effects we analyze the correlations in such a way that
the two points are taken symmetrically with respect to the center
of the system~\footnote{The two points of
    $\langle \delta n_i^{\mathrm{pol}} \delta n_j^{\mathrm{pol}} \rangle$,
    with $\vert i - j \vert = r$, have been chosen such that
    $i = (L - r + 1)/2, j = (L + r + 1)/2$ for odd r,
    and $i = (L - r)/2, j = (L + r)/2$ for even r
    (e.g. for $L = 100$ sites, $r = 1$ corresponds to $i = 50, j = 51$;
    $r = 2$ corresponds to $i = 49, j = 51$; $r = 3$ to $i = 49, j = 52$, and so on)}.
The left panel of Fig.~\ref{fig4} shows how differently such polariton correlations
behave when the system goes from the MI to DW phase, for a fixed system size.

To be more accurate, in the right panel we performed a finite-size scaling
and showed that the staggered correlation $C_{\rm DW}(r)$
approaches the zero value exponentially with $L$,
in the MI phase (a similar behavior occurs in the SF region).
On the other hand, in the DW such correlator asymptotically converges to a finite value.
In the specific, here we fix $t/\Omega=0.05$ and show that for $g_2/\Omega = 1.3, \, 1.35$
the DW order is exponentially suppressed with $L$, while for $g_2/\Omega = 1.4$
it remains finite.
The ${\mathcal O}_{\mathrm{DW}}$ order parameter reached for $L \to \infty$
is displayed in the inset as a function of $g_2$.

In order to determine the DW boundary in the phase diagram of Fig.~\ref{fig2},
we adopted the following protocol.
For a fixed value of $t/\Omega$, we start increasing $g_2$ from zero up to a finite value,
with a fixed increment $\delta g_2 = 0.05 \Omega$, and to compute the DW order parameter
for all such values of $g_2$.
The boundary of DW phase in the $g_2-t$ plane (filled circles in Fig.~\ref{fig2}),
for any fixed $t$, is located by the $g_2^*(t)$ that gives the first non-vanishing
order parameter $\mathcal{O}_{\mathrm{DW}}$.

Here we stress that, because of the arrangement
of our 1D array [see Fig.~\ref{fig1}(a)]
and of the asymmetric coupling between the atom and its right/left cavity,
the antiferromagnetic symmetry of the system is spontaneously broken.
In particular, the state $\ket{4}$ of the $L$-th atom
will be never occupied, since the transition $\ket{2} \leftrightarrow \ket{4}$
does not couple to any cavity mode [see Eq.~\eqref{hamiltonian}]. 
As a consequence, in our simulations we do not need any symmetry-breaking potential.
We can observe that the expectation value for the onsite number of polaritons
explicitly exhibits a staggered behavior, in that the occupation of the $(2n-1)$-th site
is always higher than that of the $(2n)$-th site (for any integer value of $n$).
Finally we notice that such staggering persists at finite size,
also for the set of parameters corresponding to the MI phase,
although it is extremely tiny and decreases with $L$.
This effect eventually disappears in the thermodynamic limit.

\begin{figure}[!t]
  \includegraphics[width = 0.5\hsize]{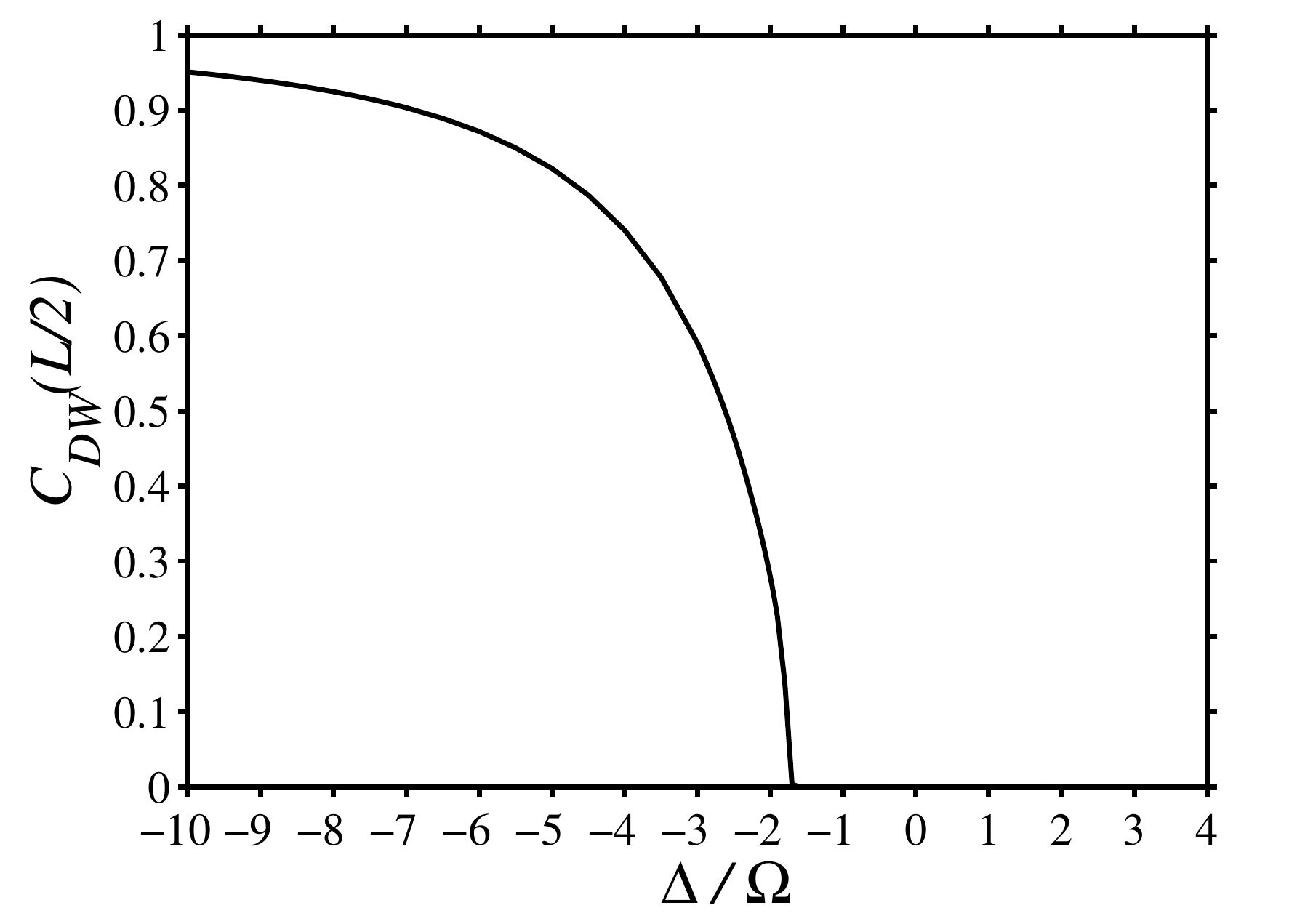}
  \caption{{\bf DW phase order parameter as a function of detuning.}
    For a positive detuning, the system will never present as a DW phase,
    meaning that $C_{DW}(L/2)=0$. Taking a negative value of the detuning,
    we observe that $C_{DW}(L/2)$ increases with $\vert \Delta \vert$.
    The parameters in this figure are $t/\Omega=0.05$, $g_1/\Omega=0.8$,
    $g_2/\Omega=1.6$ and the system size is $L=100$.}
  \label{funDelta}
\end{figure}

The extension of the DW phase depends on the cavity detuning $\Delta$.
In particular, the robustness of the order parameter increases with increasing the modulus
of the detuning (see Fig.~\ref{funDelta}). Quite remarkably, we note that a positive $\Delta$
will never stabilize an antiferromagnetic DW ordering.

\subsection{Neutral gap}
\label{sec:neutral_gap}

The analysis leading to the phase diagram in Fig.~\ref{fig2} has been corroborated
by a study of the neutral gap, which vanishes both in proximity of
the phase transitions and in the entire superfluid region.
Differently for the charge gap, it is able to detect the presence of excitations
other than particle-hole, and thus locates the boundaries of insulating regions
(as the DW) beyond the MI.

\begin{figure}[!b]
  \includegraphics[width = 0.495\hsize]{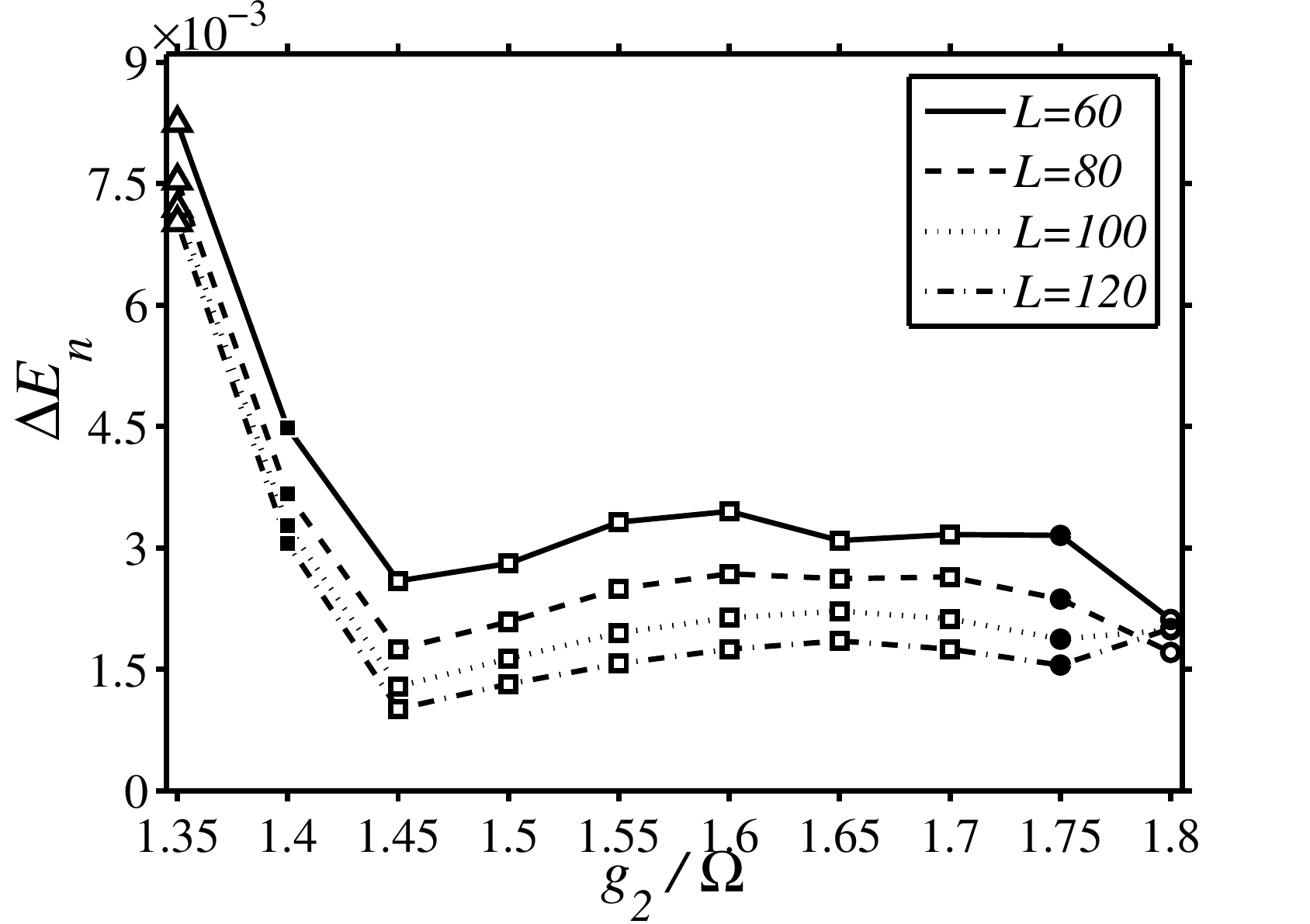}
  \includegraphics[width = 0.495\hsize]{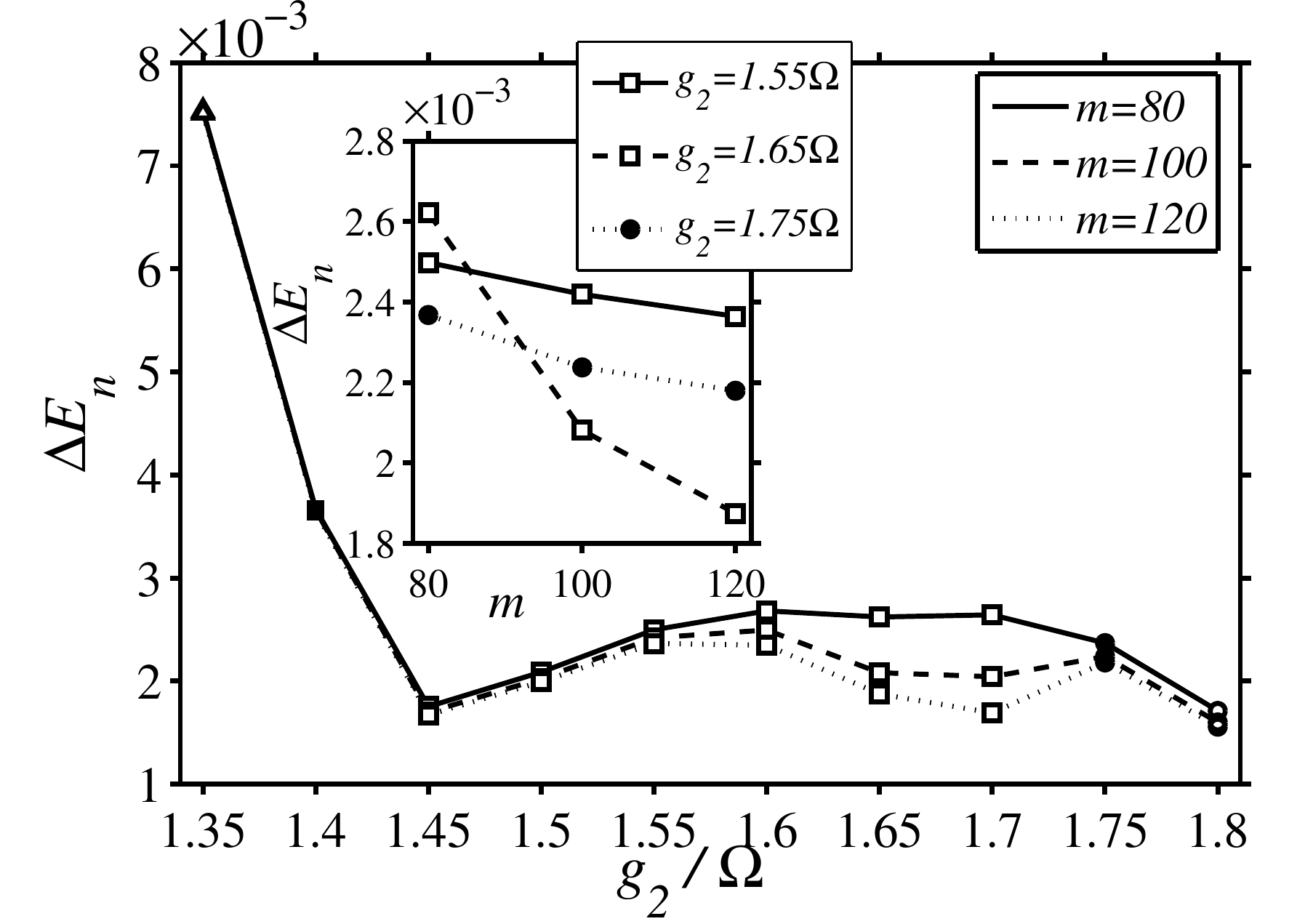}
  \caption{{\bf Analysis of the neutral gap.}
    Neutral energy gap as a function of $g_2$ for $t/\Omega=0.25$, {\it i.e.}, along
    the vertical cut depicted in Fig.~\ref{fig2}.
    In the left panel, the different curves are for various system sizes
    according to the legend, and for a fixed number of kept states $m=80$
    in the DMRG algorithm.
    The right panel evidences the convergence of the data, at fixed $L=80$, by increasing $m$
    (see also the inset, where we show the behavior of $\Delta E_n$ as
    a function of $m$, for three different values of $g_2$.}
  \label{fig5}
\end{figure}

The data displayed in Fig.~\ref{fig5} show the behavior of $\Delta E_n$ as a function
of $g_2$, for a fixed value of $t / \Omega$.
In particular we analyzed a vertical cut in the phase diagram of Fig.~\ref{fig2}
(see the rightmost vertical dotted line in that figure), where the system can be
in three different phases according to the value of $g_2$.
With increasing $g_2$, it goes from the MI phase (nonzero $\Delta E_n$, for $t/\Omega \lesssim 1.45$)
to the SF phase (zero $\Delta E_n$, for $1.45 \lesssim t/\Omega \lesssim 1.8$),
and then to the DW phase (nonzero $\Delta E_n$, for $t/\Omega \gtrsim 1.8$).

While we cannot see a clear signature of a finite gap for $g_2 = 1.8 \Omega$,
the scaling with the size displayed in the left panel of Fig.~\ref{fig5}
seems to suggest the scenario depicted above.
It is however important to stress that the DMRG simulations needed to compute the neutral gap
have to target the two lowest-lying eigenstates in a single run.
Thus they generally require a larger dimension $m$ of the effective Hilbert space,
as compared to all the other ground-state calculations discussed before.
The analysis of the neutral gap requires a careful convergence test of the
results with $m$, which we provide in the right panel of Fig.~\ref{fig5}.
We observe that the non monotonic features that are visible in the region
$1.45 \lesssim t/\Omega \lesssim 1.8$ have to be probably ascribed
to the inaccuracy of the method at small $m$ values.
This signals the presence of the gapless SF phase there, in agreement
with the results provided by the charge gap (MI/SF boundary)
and for the DW order parameter (SF/DW boundary).

\section{Summary}
\label{sec:summary}

Using the density-matrix renormalization group with open boundary conditions,
we studied the equilibrium phase diagram of a system of coupled QED cavities in one dimension.
We provided results beyond the standard model of couplings through photon hopping,
and also considered nearest-neighbor cross-Kerr nonlinearities.
Our analysis is based on a finite-size scaling of the ground-state charge
and neutral gaps, as well as of the density-wave order parameter,
for systems up to 300 sites.
We showed that, beyond the conventional Mott insulator and superfluid phases,
the presence of a nearest-neighbor nonlinear coupling can also
stabilize a density-wave ordering of polaritons.


\begin{backmatter}

\section*{Competing interests}
The authors declare that they have no competing interests.

\section*{Authors contributions}
All the authors participated in the design of the research, analysis of the results, and writing of the paper.
The DMRG code used to run all the simulations of this research has been
developed and written by D. Rossini and coworkers~\cite{dmrgB} (see also \texttt{www.dmrg.it}).
The DMRG simulations were performed by J. Jin.

\section*{Acknowledgements}
We would like to acknowledge our previous collaboration with M. Hartmann and M. Leib
which was inspiring for the present work.
This work was supported by Italian MIUR via FIRB Project RBFR12NLNA and PRIN Project 2010LLKJBX,
by EU through IP-SIQS, and by National Natural Science Foundation of China under Grant No. 11175033 and No. 11305021.



\end{backmatter}

\end{document}